\documentstyle[12pt, aasms4]{article}

\begin{document}

\title{ANISOTROPY IN THE MICROWAVE SKY:
RESULTS FROM THE FIRST FLIGHT OF BAM}
\author{G.~S. TUCKER\altaffilmark{1}, H.~P. GUSH, M. HALPERN, I. SHINKODA}
\affil{Dept. of Physics and Astronomy, University of British Columbia,
       Vancouver, BC  V6T 1Z1, Canada \\
Electronic mail: gtucker@cfa.harvard.edu}
\and
\author{W. TOWLSON}
\affil{Dept. of Physics and Astronomy, University College London,
       London WC1E 6BT, England}

\altaffiltext{1}{Current address: Harvard-Smithsonian Astrophysical
Observatory, 60 Garden Street, Cambridge, MA 02138}

\begin{abstract}
Results are reported from the first flight of a new balloon-borne
instrument, BAM (Balloon-borne Anisotropy Measurement), designed to
search for cosmic microwave background (CMB) anisotropy.  The
instrument uses a cryogenic differential Fourier transform
spectrometer to obtain data in five spectral channels whose central
frequencies lie in the range 3.7 cm$^{-1}$ to 8.5 cm$^{-1}$.  The
spectrometer is coupled to an off-axis prime focus telescope; the
combination yields difference spectra of two regions on the sky
defined by 0\fdg 7 FWHM beams separated by 3\fdg 6.  Single
differences obtained at ten sky positions show statistically
significant fluctuations. Assuming Gaussian correlated anisotropy, for
the band average 3.1 cm$^{-1}$ to 9.2 cm$^{-1}$, one finds $\Delta T/T
= 3.1^{+3.1}_{-1.1}\times 10^{-5}$ (90\% confidence
interval) for a correlation angle of 1\fdg 2. This corresponds to
$Q_{\rm flat} = 35.9^{+17.7}_{-6.3}$ $\mu$K ($1\sigma$).
\end{abstract}


\section{INTRODUCTION}

Observations of cosmic microwave background (CMB) spatial
anisotropy at intermediate angular scales provide critical information
on conditions in the early universe at redshifts $z>1000$, and put
strong constraints on cosmological scenarios.  In particular, measurements
at degree angular scales provide a sensitive test of scale free
fluctuations below the so called Doppler peak (\cite{wss}).

On an angular scale of 10\arcdeg, the Cosmic Background Explorer
(COBE) satellite has detected anisotropy of $30\pm 5$ $\mu$K
(\cite{smoot}).  Cross correlation with FIRS, an independent
observation at much shorter wavelengths (\cite{ganga}), confirms the
COBE observation and its interpretation as cosmic in origin.  White,
Scott, \& Silk (1994) provides a recent summary of measurements in
this field.  We report here the first results from a new instrument,
BAM (Balloon-borne Anisotropy Measurement), which measures at angular
scales from \slantfrac{3}{4} to several degrees, angles just larger
than typically predicted for the location the first Doppler peak.
Measurements at this scale will determine the angular spectral index
of primordial inhomogeneities.

\section{APPARATUS}

BAM consists of a cryogenic differential Fourier transform
spectrometer coupled to an off-axis prime focus telescope.  The
advantages of the BAM approach to measuring CMB anisotropy, consisting
of a clean optical system and good spectral coverage, have been
described elsewhere (\cite{halp1995}; \cite{halp1993}).  The
spectrometer (\cite{gh}) was previously used for a measurement of the
CMB spectrum (\cite{ghw}).  For this experiment, the spectrometer is
fitted with new input optics and passband-limiting filters, a new long
duration $^3$He refrigerator (Tucker et al. 1996) and is housed in a
new cryostat.  The telescope is pointed with a three-axis servo system
locked to guide stars.  The electronic and electromechanical
components of a pointing system used previously with other
balloon-borne instruments (\cite{anderegg}) were incorporated into a
new lightweight balloon gondola.

The two input collimators of the differential spectrometer view the
same portion of the aluminum primary mirror, resulting in two input
beams on the sky which are $0\fdg 70\pm 0\fdg 05$ FWHM separated by
$3\fdg 60\pm 0\fdg 05$. The optical arrangement is
sketched in Figure 1.  Particular attention is paid to the suppression
of spurious radiation entering the instrument.  A diffraction control
horn surrounds the two spectrometer entrance windows so that neither
window is visible from anywhere except in reflection off the primary
mirror.  A reflective ground shield surrounds the entire lower
portion of the instrument so that no part of the diffraction control
horn, input collimators, or primary mirror is directly illuminated by
radiation from the earth.

The interferometer has two independent bolometric detectors, one for
each output port.  Each bolometer yields a double sided interferogram
proportional to the brightness difference between the two input beams.
The scanning time for one interferogram is 0.67 s, and the optical
path length difference runs from 1.44 to $-0.91$ cm.  Spectra are
obtained {\it a posteriori} by Fourier transforming these interferograms.

\section{FLIGHT}

BAM was launched from the National Scientific Balloon Facility in
Palestine, Texas at 00:45 UT (at sunset) on 8 July 1995 and reached a
float altitude of 41.5 km at 05:00 UT.  Observations ended at 09:00
UT, in darkness, when the payload was near the end of telemetry range
and the flight was terminated.  The payload was recovered with only
minor damage.

To calibrate the instrument and measure the beam profile, three scans
of the telescope beams across the planet Jupiter were made just prior
to reaching a stable float altitude.  The scans were made in azimuth
at three fixed elevation angles, so that Jupiter passed through both
beams of the telescope.  Unfortunately, the elevation angles were not
chosen optimally so Jupiter was scanned only with the lower half of
the telescope beams. As a result the maximum possible signal from
Jupiter was never realized.  Using data about the beam shape obtained
on the ground prior to flight, along with the flight data, it is
estimated that the maximum observable signal is $1.2\pm 0.1$ times
larger than the measured response.  The data of Griffin et al. (1986)
for Jupiter have been used as an absolute reference to calibrate the
power sensitivity of the instrument.  The total calibration
uncertainty is 20\%.

Soon after reaching float altitude, an integrated circuit on board the
telescope pointing system failed.  This caused disruption of commands
controlling the telescope orientation.  The telescope remained locked
to guide stars, but slewing the telescope to new orientations could
not be carried out on the planned schedule.  After diagnosing the
problem, a switch was made to a redundant telemetry channel not
originally intended to be used for in-flight control; and a software
patch was also developed.  Thus, reliable operation of the pointing
system was re-established for the last half hour of flight.  The
instrument performed well despite these problems, but the reduction
of observing time reduced the sensitivity, and the disruption of the
planned observing pattern limited the ability to check for atmospheric
effects.

As this was the first flight of BAM, a conservative observation
strategy was adopted to allow the instrument performance to be
understood.  The telescope observes near the meridian above the North
Celestial Pole.  The left telescope beam is moved to a target spot
east of the meridian and the telescope is locked with reference to a
guide star within 5$^\circ$ of the beam.  A series of single
difference interferograms, $I^+_i$, are obtained while tracking the
guide star.  After a period of observation, called a stare, the
telescope is slewed so that the right beam is on the target spot and a
second series of interferograms, $I^-_i$, is accumulated.  The series
$I^+_i, I^-_i$ is called a wobble.  The difference signal $I^+-I^-$ is
an interferogram of the difference in brightness between the target
spot and the average of the two spots to either side of it.  Wobbling
is repeated while tracking the target spot to west of the meridian
after which a new target spot east of the meridian is acquired.  It
had been intended to initiate a new stare every three minutes, but the
command intermittency resulted in not wobbling for periods of up to
fifteen minutes in duration.

\section{DATA ANALYSIS}

The data set consists of a series of three-minute long data files
containing the bolometer and optical path length difference signals
sampled uniformly in time along with 50 channels of auxiliary data.
Each file contains approximately 263 interferograms.  About 10\% of
the data in each interferogram is deleted corresponding to the time
that the interferometer mirror is slowing and reversing direction at
the ends of the scan.  Because the bolometer signals are sampled
uniformly in time and the interferometer speed is not constant, it is
necessary to perform an interpolation of the data to acquire
interferogram samples uniformly spaced in optical delay.

These interpolated interferograms are edited as follows.  Those
interferograms corresponding to periods in which the telescope was
slewing are deleted (23.2\%).  Data contaminated by the effects of
cosmic rays striking the bolometers are removed from the remaining
data using the following algorithm.  For each stare, a mean and
variance are computed at each value of optical delay in the
interferogram.  Any individual interferogram containing a point more
than $5\sigma$ from the mean is excluded from further analysis; a
total of 6.7\% of the data are deleted this way.  To exclude smaller
cosmic ray events which are hidden in the variance caused by the
events removed above, the process is repeated on the remaining
interferograms, and a further 2.0\% of the data are deleted.  After
the above editing 61.4\% of all the data remain for analysis.  Spectra
are calculated using an FFT routine to find the Fourier transform of
each interferogram.

Interferograms obtained during ascent contain an appreciable
atmospheric signal.  These data are used to provide a reference for
phase correcting the interferograms, that is, removing the effects of
phase shifts in the bolometers and amplifiers.  Thus, optical signals
appear only in the Fourier cosine transform.  The sine component of
the phase corrected Fourier transform contains no optical data at any
frequency but provides a monitor of systematic errors.  Signals from
the atmosphere at float altitude are too weak to provide a phase
correction reference.  An independent and consistent phase correction
is obtained from the Jupiter data.

We had intended to analyze the data using a double difference
technique taking data from temporally adjacent stares in the $+$ and
$-$ positions.  However, the unreliable wobbling achieved during
flight precludes such an analysis.  Therefore only single differences
are analyzed.  Each stare is divided into two half-stares.  For each
half-stare the difference spectrum $\Delta^\nu_i = {\cal F}(I_i -
I_{\rm off}$) is calculated where the $+$ and $-$ subscripts on the
symbol $I$ have been dropped since only single differences are being
considered.  The symbol $\cal F$ indicates a Fourier transform, and
$I_{\rm off}$ is a known constant offset interferogram due to
radiation internal to the instrument.  The magnitude of the offset is
about 20 mK.  The variances of the $\Delta^\nu_i$'s,
$\sigma^2_{\Delta^\nu}$, are also computed and are an error estimate
unbiased by sky signal; these error estimates are propagated through
the remainder of the analysis.

During the flight, a gap in the anisotropy data of about an hour
occurred while the software patch was being developed.  During this
time the telescope observed another part of the sky, at a different
elevation angle than for the data reported here.  (Unfortunately,
because wobbling was absent conclusions about anisotropies could not
be drawn at this different elevation angle; the data were used,
however, to help check for systematic errors.)  The data taken before
and after the gap are analyzed separately but identically.  For each
frequency channel the data are simultaneously fitted to a model of the
sky, a constant offset, a linear drift in time, payload altitude, and
the temperature of the torque motor which drives the spectrometer scan
mechanism.  The fits are done independently for the cosine and sine
transforms and a full covariance matrix is propagated.  The reduced
$\chi^2$ for the fits range between 0.37 and 1.20.  In the worst
frequency channel the total drift during the flight is 2.5 mK while in
the best channel the drift is 500 $\mu$K.

The torque motor just referred to is located in an ambient temperature
region of the cryostat vacuum space.  Approximately every 30 minutes by
operator choice, a heater on the torque motor was turned on or off in
order to maintain the motor temperature in a useful range.  This power
variation caused the motor temperature to vary by 4 K during
the flight.  Apparently heat conducted from the motor through the
shaft to the mirror scanning mechanism differentially heated the
interferometer housing.  A significant drift (up to 600 $\mu$K)
correlated with the torque motor temperature is observed in the sine
transforms but is not apparent in the cosine transforms.  Since light
emitted from the interferometer walls is modulated differently from
light entering the interferometer through the input collimators, it is
plausible that such a signal would only appear in one phase of the
transforms, as observed.

In addition to the torque motor temperature, a number of other
temperatures and differential temperatures are measured both inside
the cryostat and elsewhere on the gondola.  With the exception of the
torque motor temperature, no correlation between $\Delta^\nu$ and any
temperature or differential temperature is observed.  In particular, a
2 K temperature gradient was induced across the primary mirror near
the end of the flight by heating one side of the mirror.  No
correlation is observed between $\Delta^\nu$ and the temperature of
the primary mirror or temperature gradient across the mirror.

Due to the limited integration time and failure to obtain a sufficient
number of {\it double} differences, the sensitivity achieved in each
frequency channel is modest and inadequate to accurately constrain the
spectral index of any observed signal; therefore we present only band
average data.  The final single differences are shown in Figure 2.

If each stare is divided instead into thirds, the analysis yields
consistent results.  However with finer subdivision, the final
differences are discrepant indicating the presence of a non-stationary
noise source.  The cross elevation of the telescope oscillates with a
peak-to-peak amplitude of 0\fdg 4 and a period of 12 s.  This may be
the source of the non-stationary noise.  However, the sensitivity on
these very long time scales is low and it has been impossible to find
and remove any correlation.  Another problem which gave concern is
that there is increased noise in the detectors when the interferometer
mirror is scanning; however, there is no evidence that this has
introduced systematic effects that could mimic a detected signal.

\section{LIMITS ON CMB ANISOTROPY}

There is strong evidence that the observed signals are from optical
sources external to the instrument.  Significant power is detected
only in the cosine component of the spectrum, which contains optical
signals, while the sine component, which does not contain optical
signals, is consistent with noise.  Given the limited integration
time, we can not reliably dismiss the possibility that the signal is
from the atmosphere.  It is also difficult to rule out definitively
earthshine and moonshine entering the instrument differentially
although much care has been exercised in designing the ground shield.
There appears to be a mild correlation between the cosine and sine
single differences (consisting of a monotonic drift in offset during
the flight).  The cause of this possible correlation is unknown.
Because the power in the correlated part of the signal does not exceed
the power in the sine component, and the power in the sine component
is negligible, it can be concluded that the small correlated component
has minimal effect on the detected power in the cosine component.  The
structure in Figure 2 does not correlate with the IRAS 100 $\mu$m maps
(Wheelock et al. 1993).

Assuming that CMB anisotropy is Gaussian correlated with a correlation
angle $\theta_c$, a likelihood analysis (\cite{readhead}) has been
performed on the full frequency band data.  For a given $\theta_c$,
the 90\% confidence interval on $\Delta T/T$ is computed by
integrating the resulting likelihood curve.  The results are shown in
Figure~3.  A value $\Delta T/T = 3.1 {+ \atop -} {3.1 \atop 1.1}
\times 10^{-5}$ at a coherence angle of $\theta_c = 1\fdg 2$ is
derived; the calibration uncertainty is not included in this number.
For the sine component of the spectrum a null signal is expected.
Applying the same analysis to the sine component, a signal consistent
with zero is found; a 95\% upper limit of $\Delta T/T = 1.89\times
10^{-5}$ is found for $\theta_c = 1\fdg 2$. This is a measure of the
instrumental sensitivity.  In particular, the signal associated with
any correlation between the sine and cosine components shown in Figure
2 is not larger than the full sine amplitude, and thus is consistent
with zero and not consistent with the power detected in the cosine
component.

Using the approximations of Steinhardt (1995) and White \& Scott (1994),
the detected cosine amplitude can be expressed as
$Q_{\rm flat} = 35.9^{+17.7}_{-6.3}$ $\mu$K where the effective spherical
harmonic of the observation geometry is $\overline\ell = 74$.

\section{CONCLUSION}

Statistically significant fluctuations in the microwave sky have been
observed. However, the spectrum has not been measured with sufficient
sensitivity to attribute these fluctuations definitively to cosmic
origin. On the other hand the new results are similar to those of
others measuring anisotropies; that is, there are scattered regions of
the sky whose temperatures are unexpectedly high.  Clearly, it is
becoming urgent to establish whether these regions pertain to the
cosmic background. This new instrument, BAM, is potentially capable of
doing this in a full flight, since it measures the spectrum of the
anisotropies.

Without pointing system problems and a longer balloon flight, an
increase of 4--8 in integration time is easily achievable.  In
addition another factor of 2--4 in reduced detector noise is easily
realizable with a new mirror scanning mechanism.  We expect to adopt
a more optimal observing strategy and spectral resolution for the
next flight.

We would like to emphasize that the modest sensitivity obtained in
this flight is the result of a series of mechanical and electrical
malfunctions and is not the result of any fundamental limit of the
instrument.  In addition all of the malfunctions have readily
identifiable causes.  Preventing a recurrence of these malfunctions
provides no serious technical challenges.

\acknowledgements

We would like to acknowledge the expert flight support provided by the
United States National Scientific Balloon Facility, Palestine, Texas.
Caitlin Davis, Miranda Jackson and Chris Padwick helped prepare for
and assisted with the flight.  The undergraduate students Gill Bakker,
Jamie Borisoff, Kevin Driedger, Dan Griggs and Chris Trautman helped
build parts of the instrument.  We would like to thank Ed Wishnow for
his extremely valuable help in Palestine, Texas during his vacation.
We thank the UBC mechanical and electrical shops.  This research was
supported by the Canadian Space Agency, the Natural Science and
Engineering Research Council of Canada and the Particle Physics and
Astronomy Research Council of the United Kingdom.

\clearpage

\figcaption[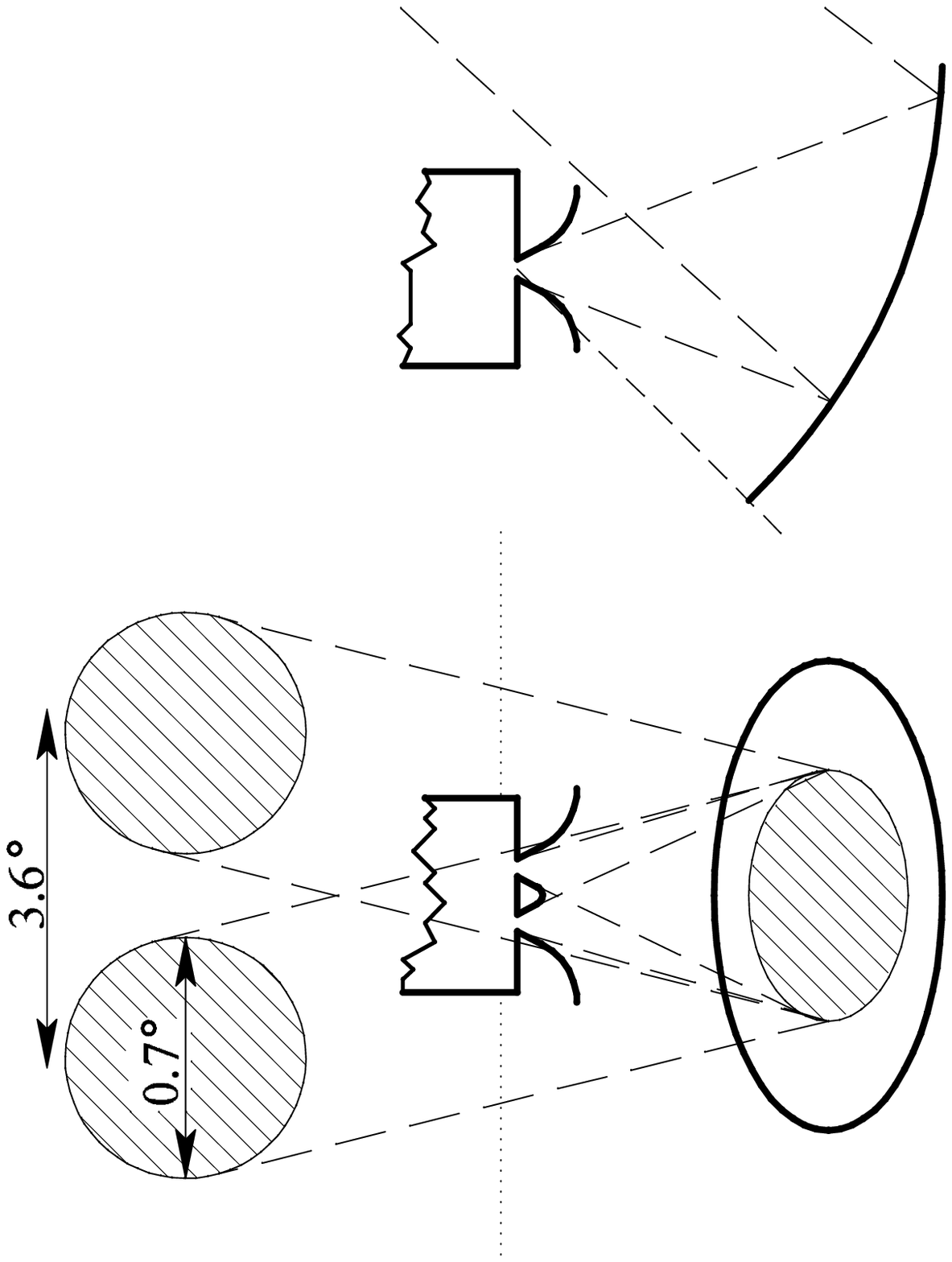]{Optical layout of the off-axis telescope.  In
the side view to the right the short dash line shows the optic axis of
the off-axis parabolic primary mirror.  The optic axis passes through
the location of the input collimators at the focus of the primary
mirror.  The long dash lines indicate the extreme rays accepted by the
input collimators.  On the left is shown the telescope as viewed from
the front.  The two input collimators are displaced laterally from the
focus and are tilted to view the same portion of the primary mirror
resulting in the beam pattern shown at the top.  The dotted line
indicates the height of the ground shield\label{fig1}}

\figcaption[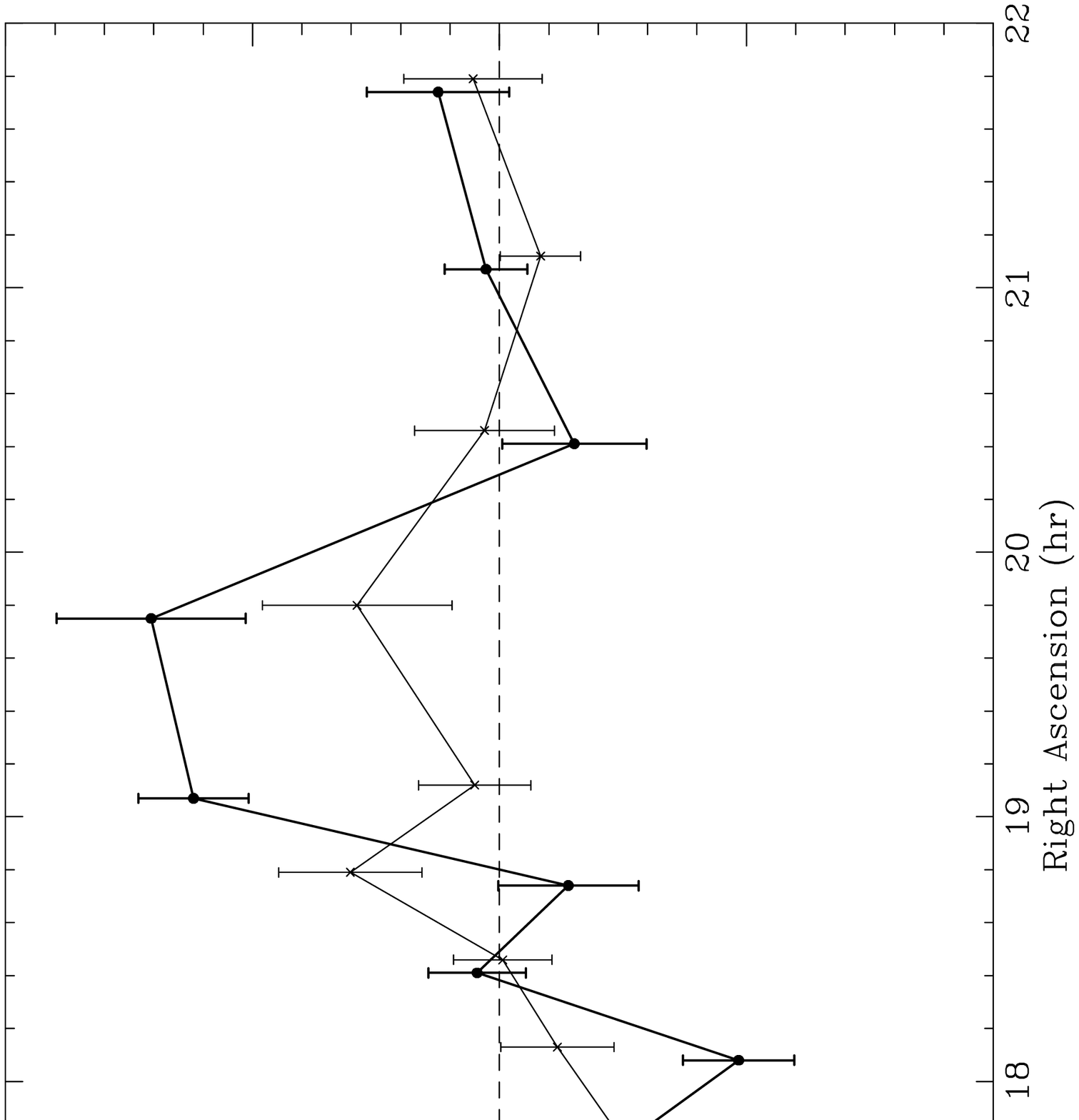]{Final single differences and statistical error
bars for the ten positions in right ascension at a declination of
about 70$^{\rm d}$.  The $\bullet$ symbols denote results for the
cosine component of the spectrum which contains the optical signal.
The $\times$ symbols indicate the sine component of the spectrum which
represents noise.  The sine symbols are offset in right ascension for
visual clarity.  There is significantly more power in the cosine than
sine component indicating that a signal has been observed. \label{fig2}}

\figcaption[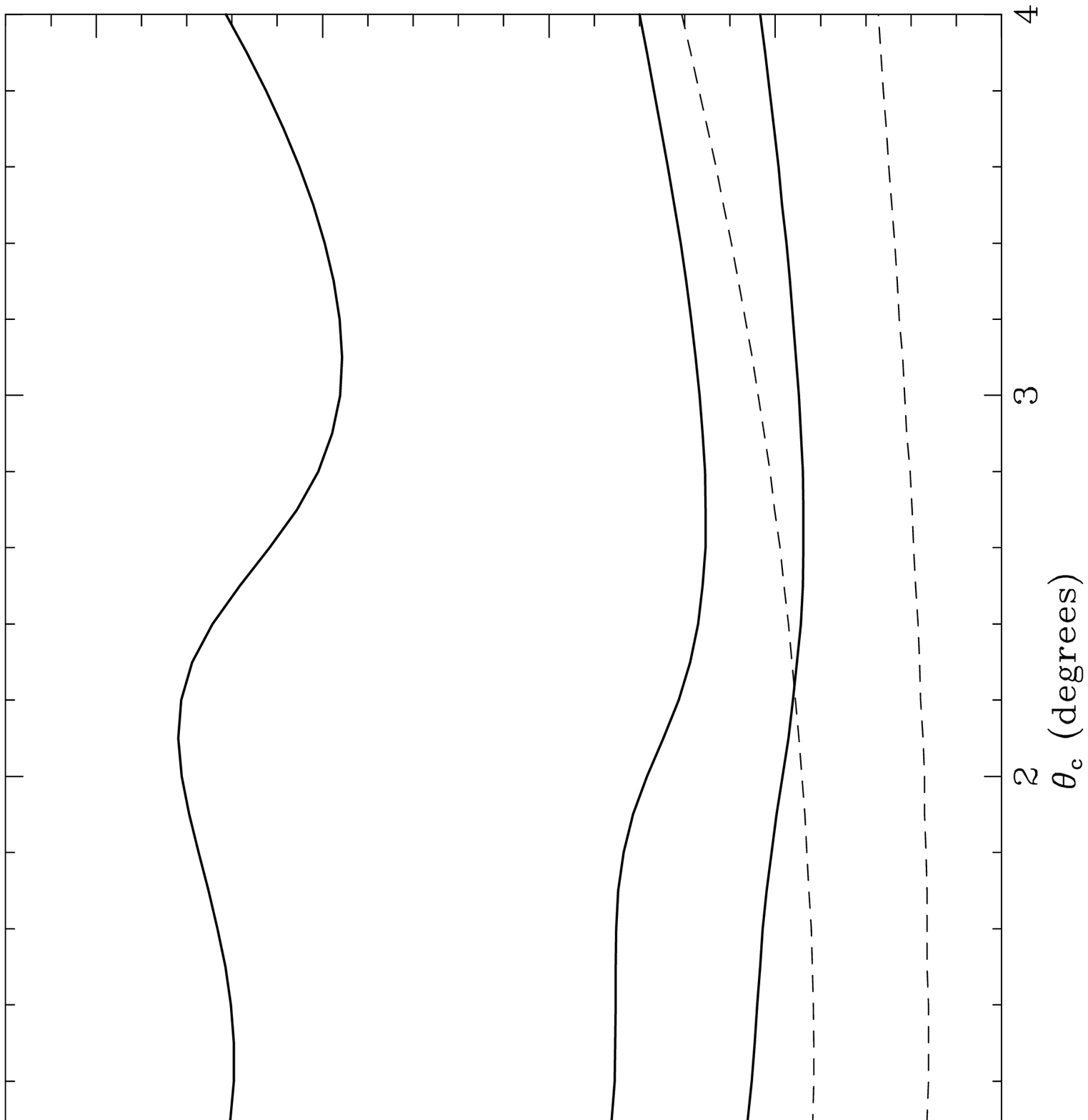]{Limits on CMB anisotropy assuming Gaussian
correlations with correlation length $\theta_C$.  The solid curves
show the peak of the maximum likelihood function and the 90\%
confidence interval for the cosine component of the spectrum.  The
dashed lines show the maximum of the likelihood function and the 95\%
upper limit on the sine component of the spectrum. Zero power can not
be ruled out for the sine component.\label{fig3}}

\clearpage

\plotone{fig1.eps}

\clearpage

\plotone{fig2.eps}

\clearpage

\plotone{fig3.eps}

\end{document}